\documentclass{agujournal2019}
\usepackage{url} 
\usepackage[inline]{trackchanges} 
\usepackage{soul}
\usepackage{amsmath}
\usepackage{amsfonts}
\draftfalse

\journalname{AGU}

\begin{document}

%
%


\title{The clock advance description of the Epidemic Type Aftershock Sequence (ETAS) model}

%
%




\authors{Matthias Holschneider\affil{1}\thanks{DFG SFB 1294}}


\affiliation{1}{University of Potsdam, Institute of Mathematics}




\correspondingauthor{Matthias Holschneider}{hols@uni-potsdam.de}



\begin{keypoints}
\item ETAS model
\item triggering of aftershocks
\item microseismicity on all scales
\item jump processes
\end{keypoints}

%
%

%
%


\begin{abstract}
In this short paper we propose to extend the ETAS model to micro-seismic
events.
For that we interpret the triggered events in an ETAS model as individual
local clock advances
of an independent background process. The solution of the ETAS model thus
becomes the sum of an infinite Markov chain of independent time adjusted
background processses.
This allows the incorporation of events at all scales. No artificial small
magnitude cutoff is needed.
\end{abstract}

\section*{Plain language summary}
A widely used model for the description of seismic activity is the ETAS
model. It describes how starting from some background activity,
earthquakes trigger new earthquakes
giving rise to a full cascade of aftershocks. Until now, it was not possible
to include microseismicity into the
description due to the infinite number of such events arising in a finite
time. In this paper I propose a
shift of interpretation of the triggering mechanism. As I show, we may
describe the triggering effect of a single
earthquake as an additional independent,  clock advanced background process. In
this formulation, the ETAS cascade
can be understood as a converging sequence of local clock advance functions
and the microseismicity can be fully incorporated
A great advantage of this concept is, that the artificial minimal triggering
magnitude can now be safely removed from
the model and net effect of the micro-seismic activities in terms of
macroscopic creep may be analyzed.

%
%

%


%
%
%
%

\section{The ETAS model}

The epidemic type aftershock model (ETAS) is usually written in terms
of time varying intensity or frequency of a Poisson point process of events
in time and
magnitude (see \cite{ogata}).
Here we shall limit ourselves to time dependent processes and do not consider
the spatial aspect
so that we have
$$
\mu(t, m | \mathcal{H}_t) = 10^{a -b m} + \sum_{t_i \leq t} h(t-t_i)
10^{\kappa + \alpha m_i + a-b m},\quad m_0\leq m < m_{max}.
$$
The notation $\mathcal{H}_t$ stands for (the $\sigma$-algebra generated by)
all events occurring before the time $t$.

The first term on the right-hand side is the background process. It is a
Poisson point process
in  time $\times$ magnitude with
intensity
$$
g(t, m) = 10^{a - bm}.
$$
The term intensity (or frequency) refers to the fact that in any time $\times$\
magnitude region $\Omega$
the number of events $N$ follows a Poisson distribution with frequency
$\Lambda$
$$
N \sim \frac{\Lambda^N}{N!} e^{-\Lambda},
$$
and the rate $\Lambda$ is simply
$$
\Lambda = \int_\Omega g(m, t) dm dt.
$$
The magnitudes are independent of the event times and are following a
Gutenberg Richter distribution law
$$
m| t_i \sim \frac{b \ln(10) 10^{- b m}}{10^{m_0} - 10^{m_{max}}},\quad
m_0\leq m\leq m_{max}.
$$
and the total Poisson rate of events with $m_0<m<m_{max}$ per time is
$$
\mbox{rate of $t_i$} = \int_{m_0}^{m_\infty} g(m) dm = \frac{1}{b
\ln(10)}10^{a}(10^{-b m_{max}} - 10^{-bm_{0}})
$$

The kernel $h$ that distributes the triggered seismicity to future times is
assumed to be normalized
$$
\int_0^\infty h(\tau) d\tau  = 1.
$$
It is commonly assumed to be of the Omori family
$$
h(t) =	\frac{C}{(1+ t/c)^p},\quad p > 1,
$$
however, for our analysis this specific form does not play any role.

Then we have to following trigger function that describes the Poisson point
process intensity of the aftershocks
generated by an event at time $\tau$ and magnitude $m^\prime$
$$
t, m| \tau,m^\prime \sim   g(m)h(t-\tau)F(m^\prime)
$$
with $F$ the productivity function
$$
F(m) = 10^{\kappa + \alpha m}.
$$

We agree, that $m(t)$ denotes the formal catalog function
$$
m(t) = m_i\ \  \mbox{if $t = t_i $},\quad
m(t) = -\infty\ \ \mbox{if no event at $t$}
$$
With this notation we have $F(m(t)) = 0$ whenever there is no event at $t$
We then can write the ETAS model more concisely as
$$
\mu(t, m | \mathcal{H}_t) = g(m) + \sum_{\tau \leq t} g(m) F(m(\tau)) h(t
- \tau)
$$
The sum over all times $\tau\leq t$ is well-defined since only a finite
number of times  actually contribute, at least as long
as $m_0>0$.

Then, the random process defined by the ETAS equation can be realized
in two different ways.

\textbf{The causal sequential picture.} We simply follow the equation as
stated. So, starting from a single
random event time $t_1$ drawn from the appropriate exponential distribution
(the inter-event time distribution of a Poisson process with
constant time rate),
we pick independently a magnitude $m_1$ from the Gutenberg Richter
distribution. This will change the Poissonian rate for all upcoming events.
We pick the next time point from the altered non-stationary Poisson rate,
sample a magnitude and continue that way.
For a finite smallest magnitude $m_0>-\infty$, this strategy is valid,
since with probability one,
in every step the non-stationary
Poisson rate of the future events is bounded, and thus the next event time
is always strictly $t_{i+1}> t_i$
yielding an increasing sequence of times.  So the ETAS model exists for some
time at least.
It may however happen, that the time intervals become shorter and shorter
leading to a run-away effect in the seismicity in finite time.
Under conditions linked to the branching ratio this however does not happen
and the sequence of \lq\lq next\rq\rq  events grows until it leaves $[0,T]$
and the whole catalog is generated
in a finite time.
This approach however fails if we want to include arbitrary small events,
 of which the frequency tends to $\infty$. Indeed then, there is no \lq\lq
 next\rq\rq\ event anymore. It is like the rational numbers,
where there isn't any \lq\lq next\rq\rq\  number after $1/2$, say.

\textbf{Generation picture.} Instead, we could also use the additivity of
the Poisson process
to proceed in a non-causal way.
The sum of two Poisson point processes with non-stationary rate $\lambda_1(t,
m)$ and $\lambda_2(t, m)$
is again a Poisson process
with rate $\lambda(t,m ) = \lambda_1(t,m)  + \lambda_2(t,m)$. This means we
can either
directly sample from $\lambda$
or we sample from $\lambda_1$ and $\lambda_2$ independently and join the
catalogs.
This leads to the following strategy. In order to obtain a random realization
of the
 ETAS model for $t \in [0, T]$
we start from a random catalog of background events defined through the
background intensity $g(t,m)$
For each of the finely many background events we sample from the triggered
events, discarding all events with $t>T$, to obtain the generation
$1$ events, and so on. With probability one, for finite $m_0$ and a branching
ratio  $<1$ this procedure comes to an end. Indeed, the total number of siblings
of the Galton-Watson process is finite. For a branching ration $>1$, the expected 
number of events triggered by each background event grows exponential with each generation.
Note however, that even for a  branching ration $>1$, still with a finite probability 
all background events in $[0, T]$ only have a finite total number of siblings and again the 
ETAS summation processes comes to an end with probability $>0$.

It has been argued, and given the above description it seems reasonable,
that the ETAS model undergoes a kind of phase transition, whenever the branching
number (i.e. the expected number of earthquakes triggered by an earthquake)
exceeds the threshold of $1$ see
e.g.~\cite{https://doi.org/10.1029/2001JB001580}.
In~\cite{PhysRevLett.126.128501} the authors analyze the criticality of the
earthquake generation process with the help of these
characteristics. For that reason, in order to allow for a sub-critical
branching ratio,
until now, the lower threshold $m_0$ below which no earthquakes are triggered
anymore had to be set to a finite positive value since otherwise,
the branching number which is based on the expected number of siblings would
be necessarily infinite.
In this paper I want to show that if instead of looking at the number of
earthquakes, we define a productivity ratio
as a measure for the aftershock activity, the ETAS model can be defined
easily to include all magnitudes down to $m_0 = -\infty$.
Clearly the number of earthquakes when including micro-seismicity to arbitrary
small magnitudes
will go to $\infty$ however the total cumulated impact on the triggering
capacity remains finite.

The lower threshold $m_0$ that is thought to be necessary for the ETAS model
to be well-defined
has to be distinguished from an observational threshold below which the
earthquakes can not be
detected anymore.
This parameter $m_0$ however constitutes an ad hoc chosen parameter for the
model. It is essentially impossible to estimate this
value from data.
It is rather unnatural to assume that the scaling law of the Gutenberg Richter
distribution is not valid down to micro seismic events.  Clearly at some
very small scale the physics may
change completely, but the assumed cut-off scales $m_0$ are way above the
limit at which in laboratory experiments the fundamental
scaling laws of earthquake physics can still be observed.

In this paper, I show, how the ETAS model formulated in the language
of jump processes and Levy processes can be extended consistently to
incorporate all scales of microseismicity.
This opens the way to build statistically useful models that absorbs the
unobserved microscale into a Brownian motion component
of some micros-seismic Levy hum. The aim of this paper however is to lay
down the basic mathematical tools from Levy type jump processes.

Although not strictly necessary mathematically we propose to describe the
triggering of earthquakes by the time advance of some background type
seismic activity. Thus, once we have understood the background seismicity
including arbitrary small magnitude events, we can understand the ETAS
process as an infinite superposition of time advanced back-ground
processes. Each of these processes has infinite many events, however the
time scales stretch in a geometric fashion making the sum of all these time
scaled background processes converge to a limit process.

We first show the principle for an ETAS model that depends only on time. In
the last chapter we show how to extend the description to a space-time
ETAS model.

\section{The ETAS model in moment space}

In~\cite{10.1046/j.1365-246x.2002.01594.x},  Kagan has proposed to write
the Gutenberg Richter law in Moment space in which case it becomes a Pareto
distribution.
Although it would be possible to do everything in the magnitude parametrization
of the event size,
we introduce an alternative parametrization of the magnitude that is better
suited for the analysis of the ETAS dynamics.
We call it the productivity moment scale $Y$ since it measures the magnitude
of an earthquake in terms of its productivity to trigger other earthquakes.
So we define
$$
Y = F(m).
$$
This is possible since $F$ is strictly monotonically growing and thus no
information is lost by replacing $m$ with $Y$.
From now on we set $m_0=-\infty$. Then the background
Poisson intensity in time $t$ $\times$ productivity moment $Y$ is stationary
and can be computed as the change of variable
formula for the density of a measure. It comes
$$
t, Y \sim \psi(Y) = \frac{g(t, F^{-1}(Y))}{F^{\prime}(F^{-1}(Y))} = \gamma
Y^{-1 - b/\alpha}, \quad 0< Y \leq Y_{max}
$$
We suppose that $\alpha > b$. This is necessary to ensure our construction
to make sense for arbitrary small moments (see below).
Since we assume the background to be stationary in time, the right-hand side
is a function of $Y$ only.
The constant $\gamma$ and the maximal productivity moment, $Y_{max}$, are given by
$$
\gamma=\frac{10^{a + \kappa b/ \alpha}}{\ln(10) \alpha} , \quad Y_{max} =
F(m_{max}) = 10^{\kappa + \alpha m_{max}}.
$$
The right hand can be understood as the symbol of a Levy jump process.
Then, for any interval $I \subset (0, \infty)$, the integral  $\int_{I}
\psi(Y)  dY$ would be the
Poisson rate of jumps with jump size in $I$.
We want to make this construction a little more explicit.
Fix an arbitrary time interval $[0, T]$.  Then consider a random catalog
generated from the Poisson intensity $\eta(t, Y) = \psi(Y)$. Since the total
integral is infinite
$$
\int_0^{T}\int_0^\infty \eta(t, Y) dt dY =  T \int_0^{Y_{max}} \psi(Y)
dY = \infty,
$$
the standard way of sampling from a non-homogeneous (in $Y$)
Poisson point process does not apply since we can not sample from a
Poison distribution $P_\lambda$ with parameter $\lambda=\infty$. However,
we may decompose the magnitudes into disjoint dyadic intervals (see
Fig~\ref{fig:levydensity})
$$
I_k = (Y_{max}2^{-{k+1}}, Y_{max}2^{-k}], \quad \bigcup_{k=0}^\infty I_k =
(0, Y_{max}].
$$
Now for each $I_k$ we may sample as usual. That is we compute the total rate
of events in $[0, T] \times I_k$
$$
N_k = \int_0^T\int_{I_k} \psi(Y) dY \sim 2^{k b/\alpha}.
$$
We then draw a Poisson variable $L_k \sim P_{N_k}$. Given this random number
of events, we distribute $L_k$ events in time and moment according to the
pdf with density
$$
p(t, Y) =  \frac{\psi(Y)}{T N_k}, \quad (t, Y) \in [0, T] \times I_k
$$
This defines a random catalog $\Xi_k$ with moments in $Y_k \in I_k$. Then,
the collection of these sub-catalogs defines the
total catalog
$$
\Xi = \bigcup_k I_k.
$$
It contains infinite many events. However, the events are still countable,
since in any slab $I_k$ we only have a finite number.
Moreover, with probability one no two events share the same time
point. Therefore, we may replace the catalog set $\Xi$ also by the
catalog function $Y(t)$ defined as follows.
It is zero except for those points $t=\tau$ at which we have an event $(\tau,
Y) \in \Xi$.  Since there is at most one, we can define
$Y(t)$ without ambiguity as the productivity moment of the event at $t$.
Note that $Y(t)$ is not a measure, it is a function that is zero almost
everywhere. However, It
may be associated with a measure via
$$
\nu_Y = \sum_\tau Y(\tau) \delta_\tau
$$
The sum is well-defined, since it actually runs only over a countable set only
and the sum of the amplitudes of the delta function have a finite value in
any finite time interval almost surely (see below).
We now consider the cumulated productivity
moments
$$
C_Y(t) =  \sum_{\tau\leq t} Y(\tau) = \chi_{[0, \infty)} \ast \nu_{Y}(t).
$$
This defines a non-decreasing function which is right continuous but not
necessarily left continuous. However, the left limits exist and thus
the  jumps are well-defined. These jumps happen at all event times
$$
Y(t) = \Delta C_Y(t) = C_Y(t) - C_Y(t^-), \quad C_Y(t^-) = \lim_{\tau\to t,
\tau < t } C_Y(\tau).
$$
This so called cadlag property (continue \`a droite limit \`a gauche) allows us
to map seismic catalogs containing arbitrary small events
to such well-defined semi-continuous non-decreasing functions.
Vice versa any such non-decreasing cadlag function defines a possibly infinite,
not time-ordered,
seismic catalog and a catalog function $Y(t)$.
As a random process, the process $C_{Y}(t)$ has independent increments and all
its finite increments are infinitely divisible (see e.g.~\cite{doob}).

\begin{figure}
\noindent\includegraphics[width=\textwidth]{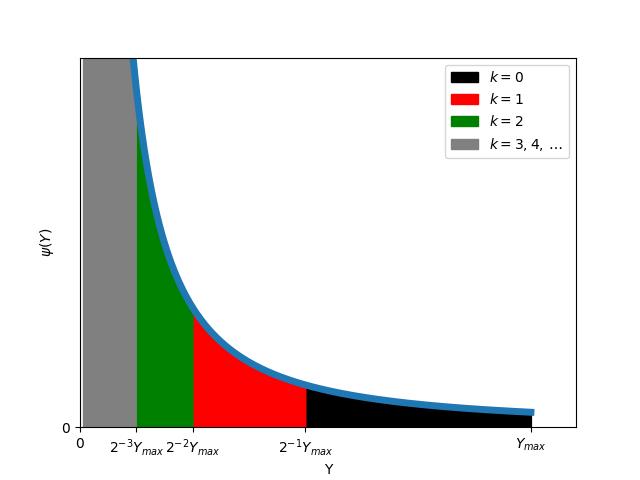}
\caption{The jump distribution can be cut into dyadic blocks.}
\label{fig:levydensity}
\end{figure}

\begin{figure}
\noindent\includegraphics[width=\textwidth]{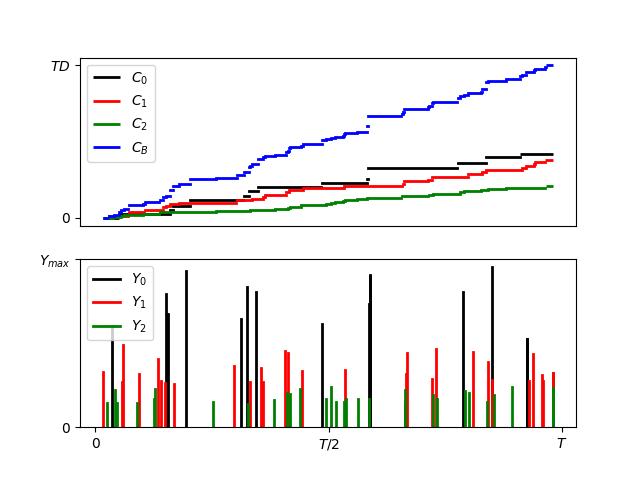}
\caption{In the upper figure is depicted a random sample the Levy path for
the first three moment blocks. The blue line is an approximation to the limit by using 
only the first $3$ blocks.}
\label{fig:levy_ranodm_dyadic}
\end{figure}

In order to actually compute (better approximate) such a function numerically
we can again resort to the decomposition $I_k$.
If $Y_k(t)$ and $C_{Y_k}(t)$ are the corresponding catalog functions and
cumulated moment functions, we then have the following scaling relation
$$
Y_{k+1}(t) \simeq 2^{ -1} Y_k(2^{\alpha / b} t),\quad C_{Y_{k+1}}(t)
\simeq 2^{-1}C_{Y_{k}}(2^{\alpha/b}  t)
$$
where the symbol $\simeq$ means both process have the same distributions.
Indeed, going from block $k$ to block $k+1$ the jump-size is scaled by $1/2$ whereas
the frequency increases by a factor $2^{\alpha/b} > 1$.
Therefore, starting from the process $Y^0$ and $C_{Y^0}$ we have the following
Weiestrass fractal function like multi-scale expansion
for the background process $Y^B$ and $C_{Y^B}$
$$
Y^{B}(t) \simeq \sum_{k=0}^\infty 2^{ - k } Y_{0}(2^{k\alpha/b} t), \quad
C_{Y^B}(t) \simeq \sum_{k=0}^\infty 2^{-k} C_{Y^0}(2^{k\alpha/b} t)
$$
The terms in the sum are each independent, rescaled (in time and event/jump
size) realizations of the common band-limited background process.
In Figure~\ref{fig:levy_ranodm_dyadic} we have shown an example of this
construction.

The average value of $C_{Y^B}(t)$ may be computed explicitly. By Campbell's
theorem (see~\ref{appendix}) we have
$$
\mathbb{E} ( C_{Y^B}(t)) = \int_0^t \int_0^{Y_{max}} Y \psi(Y) dY =
t \mathbb{E}(Y)
$$
Therefore with probability one the growth rate of $C_{Y^B}(t)$ is given by
the expectation of the productivity of the background process.
$$
D = \lim_{t\to\infty} \frac{1}{t}C_{Y^B}(t) = \mathbb{E}(Y) = \mathbb{E}(F(m))
$$
So explicitly
$$
D = \int_{Y>0} Y \psi(Y) dY = \int g(m) F(m) dm = \gamma \int_0^{Y_{max}}
Y^{-b/\alpha} =
\frac{\gamma Y_{max}^{1-b/\alpha}}{ 1- b/\alpha}
$$
or in terms of the magnitude limit $m_{max}$
$$
D = \frac{10^{a+\kappa+(\alpha-b)m_{max}}}{\ln(10)(\alpha-b)}.
$$
To ensure a finite average productivity coming from the small events, we
need to assume
$$
\alpha > b, \quad Y_{max} < \infty.
$$
This same condition is also necessary and sufficient that all the previous
computations have a well-defined meaning.
In the next section we shall add the condition $D<1$ to ensure the ETAS
model to
be well-defined.

In general a rate $\eta(t, Y)$ of jumps may be time dependent
and the band limited construction can be adapted accordingly.
For the ETAS model a simplification however occurs in case that the
distribution
of triggered event magnitudes does not depend on the magnitude of the
triggering event.
That is, we assume for simplicity that Bath's law does not hold. Then the
triggered seismicity may be absorbed in a change of time as we shall see now.
In order to include Bath's law, we would have to include additional filter operators,
which we leave for now. 

\section{The clock advance picture}
\label{sec:clockonedimen}

We now rewrite the triggering process using the language of Levy pure jump
processes.
For a single triggering event at $\tau$ and moment $Y(\tau)$ the next
generation events triggered by
this event are drawn from
the rate
$$
Y, t| \tau Y(\tau) \sim  \eta(t, Y(\tau)) = \psi(Y) Y(\tau) h(t-\tau).
$$
We now use the following fact that shows how amplitude modulation of the
background process can be
absorbed into time deformations.

{\bf Fact} If a jump process has non-stationary jump size Poisson intensity
$\eta(t, Y)$ that factorizes
$$
\eta(t, Y) = f(t) \psi(Y)
$$
then the Levy process of the cumulated $Y$ satisfies
$$
C_Y^\eta(t)  = C_Y( F(t)), \quad F^\prime(t)  = f(t),\ \ F(0) = 0
$$
The equality means, the right and left side have the same distribution.
This is intuitively clear since if locally the time is scaled by $F^\prime=f$
its local
frequency is multiplied by the
same factor. 

We introduce the elementary clock advance function $H$
$$
H(t) = \int_0^t h(\tau) \, d\tau.
$$
Then let $C_{Y^B}(t)$ be an independent process of cumulated background
moments.
The triggered events by the event at $\tau, Y(\tau)$ can therefore be
identified with the
clock modified background process
$$
C_{Y^B}( Y(\tau) H(t-\tau)),
$$
or on the catalog function itself
$$
Y^B( Y(\tau) H(t- \tau))
$$
If we have two events at $\tau_1$ and $\tau_2$ we have for the catalog
$$
Y^{B,1}( Y(\tau_1) H(t - \tau_1)) + Y^{B,2}(Y(\tau_2) H(t- \tau_2))
$$
with $Y^{B,i}$, $i=1,2$ two independent realizations of the background.
Since the sum of two independent Poisson processes with two rates is the
same as a single process with
the sum of the rates we can the above process by time deforming a single
realization of the background
$$
Y^B(  Y(\tau_1) H(t - \tau_1) + Y(\tau_2) H(t- \tau_2)).
$$
We therefore can define a total clock transform function that is applied to
the background process
via
$$
H^\ast = \sum_{\tau} Y^B(\tau) H(t- \tau) = H \ast \sum_\tau
Y^B(\tau)\delta_\tau = h \ast C_{Y^B}(t) =
H\ast \nu_{Y^B}(t).
$$
The process of the first generation of triggered events is then
$$
C_{Y^{T}}(t) = C_{Y^B}(H^\ast(t)).
$$
Note that the local clock function $H^\ast$ inherits the continuity from $h$,
and it is
strictly monotonically growing. It therefore and defines a valid time.

\begin{figure}
\noindent\includegraphics[width=0.9\textwidth]{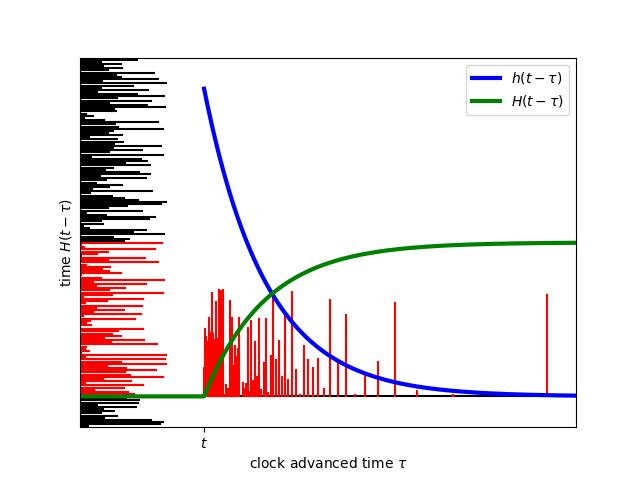}
\caption{The triggered events can be realized by modifying the local time
of an independent background process. Here we illustrate this by an event of unit size $Y=1$.
The clock modified catalog is displayed horizontally whereas the original background catalog is
along the vertical axis. Only the events colored in read will be in the locally-clock-advanced 
catalog.
}
\label{fig:levydensity}
\end{figure}

This can be iterated to obtain the sequence of triggered aftershocks. If
$Y^{T, k}(t)$ is the catalog function of the
triggered events triggered at generation $k$, we define the clock advance
function as
$$
H^{*, k}(t) = h \ast C_{{Y}^{T, k}}(t) = H \ast \nu_{Y^{T,k}}(t)
$$
and the triggered events and the corresponding Levy path at generation $k+1$
are then
$$
Y^{T, k+1}(t) = Y^{B, k}( H^{\ast, k}(t)),\quad C_{Y^{T, k+1}} = C_{Y^{B,
k}}( H^{\ast, k}(t))
$$
with independent
background process $Y^{B, k}(t)$.
The ETAS catalog function / path would then be
$$
Y^{ETAS}(t) = Y^{B}(t) + \sum_{k=1}^\infty  Y^{T, k}(t), \quad
C_{Y^{ETAS}}(t) = C_{Y^{B}}(t) + \sum_{k=1}^\infty  C_{Y^{T, k}}(t).
$$
Observe that the cumulated productivity moments on the right are not a Levy
process, since the
increments are not independent.
However, they constitute still a non-decreasing, cadlag (continue \`a droite,
limit \`a gauche)
function. The jumps are precisely the events of the infinite catalog
$$
Y^{ETAS}(t) = \Delta C_{{Y}^{ETAS}}(t).
$$

We still have to show, that the sum of random functions actually converges
almost-surely.
This will be the case whenever the expected
productivity is $<1$
$$
D = \mathbb{E}(Y) < 1 \Leftrightarrow \mbox{ETAS sum converges (with
probability one)}
$$
We only give a schematic proof here.
For that note that the longtime slope of the background Levy path, $C_{Y^B}(t)$
is $D$, $C_{Y^B}(t) / t \to D$ for $t\to\infty$.
Suppose $C_Y^{k}(t)$ has a slope of $D_{k}$
$$
t^{-1} C_Y^{k}(t) \to D_k\quad t\to\infty
$$
Then the clock advance function $H^{\ast, k}(t)=h\ast C_Y^k(t)$
has the same slope since convolution with $h$ reproduces the slope
$$
\int_0^t h(t-\tau) \tau d\tau = -H(t-\tau)\tau \vert_0^t + \int_0^{t} H(\tau
-t) d\tau  = \int_0^t H(\tau) d\tau
 \sim t\quad (t\gg 1)
$$
Therefore the time in generation $k$ is asymptotically rescaled by $D_k$. If
this is applied to
the background it produces a process with
slope
$$
t^{-1} C_{Y^B}( H^{\ast, k}(t)) \to D  D_k\quad \Rightarrow D_{k+1} = D D_k
= D^{k+1}.
$$
By induction, we have a geometric series which can be summed to show that
$$
\lim_{t\to\infty} \frac{1}{t} C_{Y^{ETAS}}(t) = \frac{D}{1-D}
$$
Note however that this is a rather hand-wavy non-rigorous argument since it
does not take into
account the random nature of these slopes.
We refer to a forthcoming paper on this subject where we use a martingale
argument.


\section{A Markov chain in infinite dimensional state space}

It is interesting to note, that our algorithm can be cast into a Markov
process with an infinite demensional state space.
Consider a sequence $C_B^{k}(t)$,  of independent background processes. The
state is given by a pair of
$(E(t), W(t))$ of non-decreasing right-continuous functions. The transition
process is defined through
$$
E_{k+1}(t) = E_k(t) + W_k(t), \quad W_{k+1} = C_B^{k+1}( h * W_k(t))
$$
Then, starting from $(0, C_B^{0})$ this defines a Markov chain. Under the
stability condition above we have
that the ETAS sampling paths are obtained as a limit
$$
\lim_{k\to\infty} E_k = Y^{ETAS}
$$

\section{Extension to spatial processes}

On the level of Poisson intensity in time and trigger moment space the
triggering dynamic can be written as
$$
\eta^{k+1}(t, Y) = \sum_{\tau} Y^{T, k}(\tau) h(t-\tau) \psi(Y)
$$
Sampling from this Poisson density can be achieved through time deforming a
sample from an independent background sample $Y^{B, k}(\tau)$
$$
Y^{T, k}(t) = Y^{B, k}(H^{\ast, k}(t))
$$
This description can be extended to spatially distributed ETAS processes.

For this we need to consider space-time Point processes described through
an intensity
$\eta(x, t, Y)$ of the form
$$
\eta(x, t, Y) =  f(x, t) \psi(Y).
$$
This factorization comes from our assumption, that the magnitude distribution
and hence the
production moment distribution of the triggered events is the same as the
one of the background.

Instead of considering Poisson point processes in the space $\times$ time
$\times$ moment space,
we can again look at the cumulated moments. For that for a spatial domain
$\Omega$
the cumulated $Y$ of the events taking	place in $\Omega$ produce a Levy
jump process.
$$
C_{\Omega, Y}(t) = \sum_{(x, \tau)| x \in \Omega, \tau\leq t} Y(\tau, x)
$$
It can be realized through a clock advance of the background process.
For this we define a spacial clock advance function of every space point
$x$ via
$$
H(x, t) = \int_0^t f(x, \tau) d\tau.
$$
Then, the clock advance in $\Omega$ is
$$
H_\Omega(t) = \int_\Omega H(x, t) dx = \int_0^t \int_\Omega f(x, \tau) dx d\tau
$$
The process of the time points of the events in $\Omega$ as
described through $\eta$
can be realized by clock modifying the one dimensional
background process $C^B_Y(t)$ of section~\ref{sec:clockonedimen}.
$$
C_{\Omega, Y}(t) \sim C_{Y^B}( H_\Omega( t ))
$$
On the level of the catalog function over space $\times$ time itself we
can write
for the clock advanced catalog
$$
Y^\eta(x, t) = Y^B(x, H(x, t)).
$$
We are actually considering here a Levy process valued measures (see
e.g~\cite{Griffiths_2021}).
In the spatial extension of the ETAS process, the seismicity is redistributed
not
only in time but also in space via a kernel, which may depend on $Y$. The
conditional
Poisson intensity reads
\begin{equation}
      \label{eq:kernel}
x, t, Y| x^\prime, \tau, Y^\prime \sim
\Gamma(x | x^\prime, Y^\prime) \psi(Y) Y^\prime h(t - \tau)
\end{equation}
We suppose that $\Gamma$ is normalized
\begin{equation}
\label{eq:gammaloc}
\int \Gamma(x| \dots )dx = 1
\end{equation}
so that $Y$ maintains its meaning as productivity.
Then the ETAS equations read
$$
\mu(x,t,Y| \mathcal{H}_t) = \psi(Y) +
\sum_{(x^\prime, \tau)| \tau \leq t} \Gamma(x| y^\prime, Y(x^\prime, \tau))
Y(x^\prime, \tau) h(t-\tau) \psi(Y)
$$
The ETAS generation iteration is now defined as follows.
Let $B^k(t, x)$, $k=0, 1, \dots$ be independent
background processes. We then consider the following Markov chain of pairs
of processes $(E, W)$ starting from $(0, B^0)$
$$
E_{k+1}(x,t) = E_{k}(x,t) + W_k(x,t),
\quad W_{k+1}(x, t) = B^{k+1}(x, H_{k}(x, t))
$$
The Poissonian intensity of generation $k+1$ can be computed from the events
at generation $k$
as
$$
\eta^{k+1}(x,t,Y) = f^{k+1}(x,t) \psi(Y),
$$
with
$$
f^{k+1} (x, t) =\sum_{x^\prime, \tau} \Gamma(x| x^\prime, W_k(x^\prime,\tau))
W_k(x^\prime, \tau) h(t-\tau)
$$
The clock advance function that maps background events to triggered events
reads
$$
H_k(x, t) =\sum_{x^\prime, \tau} \Gamma(x| x^\prime, W_k(x^\prime, \tau))
W_k(x^\prime, \tau) H(t-\tau)
$$
If this process converges, the resulting limit process of $E_k$, $k\to\infty$
is the ETAS process including microseismicity.
Again, the condition for convergence is that the expected productivity is $<1$.

\section{Conclusion and outlook}

In this paper I have shown that the concept of Levy jump processes allows
the extension
of the ETAS model to arbitrary small earthquakes under the condition that the
expected productivity remains  $<1$. The productivity is measured in units
of the
background process and not in terms of the number of siblings of a triggering
earthquake. The
latter is infinite all the time if we include micro seismicity.
This shows that the nonphysical small magnitude cutoff can be removed in a
consistent way.
The consequences for the estimation of the ETAS parameters needs to be
analyzed.
It is clear, that the specific form of the Gutenberg Richter law does not
play a role.
We only need that the productivity function has a first moment. Therefore,
for instance we may assume
a completely scale-free Gutenberg Richter law $\sim 10^{-bm}$ without cutoff
but with a two-scale productivity law
$$
F(m) \sim 10^{\alpha_- m},\quad m<m^\ast,\quad \mbox{and}\quad F(m) \sim
10^{\alpha_+ m}\quad m\geq m^\ast
$$
We only need
$$
\alpha_- > b,\quad m<m^\ast\quad \mbox{and}\quad \alpha_+ < b \quad m>m^\ast,
$$
to ensure a finite total productivity function. Adjusting the coupling
constants we can achieve a
mean moment productivity $<1$, which ensures the convergence of the ETAS
iteration.

\acknowledgments

This work was financed through the SFB1294 CRC research grant.

\appendix
\section{Campbell's theorem}
\label{appendix}

We need the theorem of total expectation
$$
\mathbb{E}(A) = \mathbb{E}_B\mathbb{E}_A(A|B),
$$
Moreover we need Campbell's theorem. It states that
for a point process with intensity $\Lambda$ over some domain and a function
$g$ the
random sum
$$
G = \sum_x g(x)
$$
has mean value
$$
\mathbb{E}(G) = \int g(x) \Lambda(x) dx
$$
Therefore, if $\Lambda$ itself is a random process, we may apply the iterated
expectation and it comes
$$
\mathbb{E}(G) =  \int g(x) \mathbb{E}(\Lambda(x)) dx
$$

\bibliography{agusample}

\begin{thebibliography}{}

\bibitem [\protect \citeauthoryear {%
Doob%
}{%
Doob%
}{%
{\protect \APACyear {1953}}%
}]{%
doob}
\APACinsertmetastar {%
doob}%
\begin{APACrefauthors}%
Doob, J\BPBI L.%
\end{APACrefauthors}%
\unskip\
\newblock
\APACrefYear{1953}.
\newblock
\APACrefbtitle {Stochastic processes} {Stochastic processes}.
\newblock
\APACaddressPublisher{New York}{John Wiley \& Sons}.
\newblock
\APACrefnote{MR 15,445b. Zbl 0053.26802.}
\PrintBackRefs{\CurrentBib}

\bibitem [\protect \citeauthoryear {%
Griffiths%
\ \BBA {} Riedle%
}{%
Griffiths%
\ \BBA {} Riedle%
}{%
{\protect \APACyear {2021}}%
}]{%
Griffiths_2021}
\APACinsertmetastar {%
Griffiths_2021}%
\begin{APACrefauthors}%
Griffiths, M.%
\BCBT {}\ \BBA {} Riedle, M.%
\end{APACrefauthors}%
\unskip\
\newblock
\APACrefYearMonthDay{2021}{{\APACmonth{05}}}{}.
\newblock
{\BBOQ}\APACrefatitle {Modelling Lévy space‐time white noises} {Modelling
  lévy space‐time white noises}.{\BBCQ}
\newblock
\APACjournalVolNumPages{Journal of the London Mathematical
  Society}{104}{3}{1452–1474}.
\newblock
\begin{APACrefDOI} \doi{10.1112/jlms.12465} \end{APACrefDOI}
\PrintBackRefs{\CurrentBib}

\bibitem [\protect \citeauthoryear {%
Helmstetter%
\ \BBA {} Sornette%
}{%
Helmstetter%
\ \BBA {} Sornette%
}{%
{\protect \APACyear {2002}}%
}]{%
https://doi.org/10.1029/2001JB001580}
\APACinsertmetastar {%
https://doi.org/10.1029/2001JB001580}%
\begin{APACrefauthors}%
Helmstetter, A.%
\BCBT {}\ \BBA {} Sornette, D.%
\end{APACrefauthors}%
\unskip\
\newblock
\APACrefYearMonthDay{2002}{}{}.
\newblock
{\BBOQ}\APACrefatitle {Subcritical and supercritical regimes in epidemic models
  of earthquake aftershocks} {Subcritical and supercritical regimes in epidemic
  models of earthquake aftershocks}.{\BBCQ}
\newblock
\APACjournalVolNumPages{Journal of Geophysical Research: Solid
  Earth}{107}{B10}{ESE 10-1-ESE 10-21}.
\PrintBackRefs{\CurrentBib}

\bibitem [\protect \citeauthoryear {%
Kagan%
}{%
Kagan%
}{%
{\protect \APACyear {2002}}%
}]{%
10.1046/j.1365-246x.2002.01594.x}
\APACinsertmetastar {%
10.1046/j.1365-246x.2002.01594.x}%
\begin{APACrefauthors}%
Kagan, Y\BPBI Y.%
\end{APACrefauthors}%
\unskip\
\newblock
\APACrefYearMonthDay{2002}{03}{}.
\newblock
{\BBOQ}\APACrefatitle {Seismic moment distribution revisited: I. Statistical
  results} {Seismic moment distribution revisited: I. statistical
  results}.{\BBCQ}
\newblock
\APACjournalVolNumPages{Geophysical Journal International}{148}{3}{520-541}.
\newblock
\begin{APACrefURL} \url{https://doi.org/10.1046/j.1365-246x.2002.01594.x}
  \end{APACrefURL}
\newblock
\begin{APACrefDOI} \doi{10.1046/j.1365-246x.2002.01594.x} \end{APACrefDOI}
\PrintBackRefs{\CurrentBib}

\bibitem [\protect \citeauthoryear {%
Nandan%
, Ram%
, Ouillon%
\BCBL {}\ \BBA {} Sornette%
}{%
Nandan%
\ \protect \BOthers {.}}{%
{\protect \APACyear {2021}}%
}]{%
PhysRevLett.126.128501}
\APACinsertmetastar {%
PhysRevLett.126.128501}%
\begin{APACrefauthors}%
Nandan, S.%
, Ram, S\BPBI K.%
, Ouillon, G.%
\BCBL {}\ \BBA {} Sornette, D.%
\end{APACrefauthors}%
\unskip\
\newblock
\APACrefYearMonthDay{2021}{Mar}{}.
\newblock
{\BBOQ}\APACrefatitle {Is Seismicity Operating at a Critical Point?} {Is
  seismicity operating at a critical point?}{\BBCQ}
\newblock
\APACjournalVolNumPages{Phys. Rev. Lett.}{126}{}{128501}.
\PrintBackRefs{\CurrentBib}

\bibitem [\protect \citeauthoryear {%
Ogata%
}{%
Ogata%
}{%
{\protect \APACyear {1988}}%
}]{%
ogata}
\APACinsertmetastar {%
ogata}%
\begin{APACrefauthors}%
Ogata, Y.%
\end{APACrefauthors}%
\unskip\
\newblock
\APACrefYearMonthDay{1988}{}{}.
\newblock

\newblock
\APACjournalVolNumPages{Journal of the American Statistical
  Association}{83}{401}{9-27}.
\PrintBackRefs{\CurrentBib}

\end{thebibliography}

\end{document}